\documentclass[11pt]{llncs}
\usepackage{color}

\newcommand{\defeq}{\stackrel{\Delta}{=}}

\newcommand{\ket}[1]{| #1 \rangle}

\newcommand{\tensor}{\otimes}

\begin{document}
\title{Quantum search for multiple items using parallel queries}

\author{
Lov K.\ Grover\inst{1} \and Jaikumar Radhakrishnan\inst{2} }

\authorrunning{Grover and Radhakrishnan}

\tocauthor{Lov K.\ Grover (Bell Labs) and
Jaikumar Radhakrishnan (Tata Institute of Fundamental Research)
}

\institute{
Bell Laboratories,
Lucent Technologies,\\
600--700 Mountain Avenue, Murray Hill, NJ 07974\\
e-mail: lkgrover@bell-labs.com
\and
School of Technology and Computer Science,\\
Tata Institute of Fundamental Research, Mumbai 400005, India\\
e-mail:jaikumar@tifr.res.in
}

\maketitle

\begin{abstract}
In the quantum database search problem we are required to search for
an item in a database. In this paper, we consider a generalization of
this problem, where we are provided $d$ identical copies of a database
each with $N$ items which we can query in parallel.  Then, given $k$ items,
we are required to determine the locations where these items are
stored.

We show that any quantum algorithm for this task must perform
$\Omega\left( \sqrt{\frac{Nk}{d \min\{d,k\}}} \right)$ parallel
queries. We also design a simple algorithm whose performance comes
within a factor $O(\log d)$ of this lower bound.

Our lower bound can be considered to be a generalization of a result
of Zalka~\cite{Zalka} who considered the case $k=1$ and $d$ arbitrary.
Our upper bound can be considered to be a generalization of the
following two results: first, a result of Boyer, Brassard, Hoyer and
Tapp~\cite{BBHT} who showed how to search for one of $k$ items in a
database, and second, a result of Heiligman~\cite{Heiligman}, which
showed that $O(\sqrt{Nk})$ queries suffice for locating all items.
\end{abstract}

\section{Introduction}

In the database search problem, we are given an item and are required
to find the location where it is stored. The goal is to perform the
task making as few queries to the database as possible. The quantum
database search algorithm of Grover~\cite{Grover} shows that this task
can be performed with $O(\sqrt{N})$ queries to the database, where $N$
is the size of the database. On the other hand it is well known that
no classical (randomized) algorithm can perform this task with less
than $\Omega(N)$ queries.

Zalka~\cite{Zalka} showed that Grover's algorithm for database search
is optimal. Furthermore, he showed that if we are provided access to 
multiple databases, and allowed to query them in parallel, then the
quantum algorithm that divides all locations equally among the
databases and performs parallel independent searches cannot be
improved. Thus, quantum search algorithms for searching one item 
using parallel queries is well understood.

The other extreme, when there are multiple items to be searched using
queries to just one database has also been studied. Boyer, Brassard,
Hoyer and Tapp ~\cite{BBHT} showed that if it is known in advance that
exactly $k$ of the items are present in the database, then using
$O(\sqrt{\frac{N}{k}})$ queries, one can locate of one of these
items. Building on this, Mark Heiligman~\cite{Heiligman} observed that
one can determine the locations of all $k$ items in time
$O(\sqrt{Nk})$.

The general problem of searching for multiple items using parallel
queries does not seem to have been addressed before. In this paper, we
consider the following problem (see below for a more formal
statement).  We are given $d \leq \sqrt{N}$ copies of a database of
$N$ items. We are given $k \leq \sqrt{N}$ items and promised that the
database contains all of them.  We are required to find the locations
of all these items.  We derive a lower bound for this general problem
and show an algorithm with performance close to this lower bound. Let
$q(N,d,k)$ be the minimum number of parallel queries made by any
quantum algorithm for the above task. Our results imply that there
exists constants $c_1>0$ and $c_2>0$ such that for all large $N$
$$ \sqrt{\frac{Nk}{d\min\{k,d\}}} \leq q(N,d,k) \leq \sqrt{\frac{Nk\lg
      d}{d\min\{k,d\}}}. $$  
In fact, for certain ranges of $k$ and $d$, the upper bound and lower
bound differ only by a constant factor.  See Theorem~\ref{thm:ub}
below for the precise statements of our upper bound.

\subsection{Background and notation}

We assume that the reader is familiar with the basics of quantum
circuits, especially the quantum database search algorithm of
Grover~\cite{Grover} (see, for example, Nielsen and
Chuang~\cite[Chapter 6]{NC}). 

\paragraph{Database search:}  The database is modeled as a function
$f: \{0,1\}^n \rightarrow \{0,1\}^m$. The elements of $\{0,1\}^n$ will
be referred to as addresses, and will be identified this set with
$[N]\defeq \{0,1,2,\ldots, N-1\}$, where $N=2^n$. We refer to elements
of $\{0,1\}^m$ as items. We say that $y \in \{0,1\}^m$ is in the
database $f$ if there is some $x\in \{0,1\}^n$ such that $f(x)=y$.  In
the quantum model, this database is provided to us by means of an
oracle unitary transformation $T_f$ which acts on an $(n+m)$-qubit
space by sending the basis vector $\ket{x}\ket{z}$ to $\ket{x} \ket{z
\oplus f(x)}$, where $x\in \{0,1\}^n$ and $z\in \{0,1\}^m$. Our
database search can then be formulated as follows.

\begin{figure}
\framebox{\parbox[t]{5.5in}{
\begin{description}
\item[Input:] We are allowed access to $d$ copies of the database $f$,
via the unitary transformation
\[ T^{\tensor n}_f: \ket{x_1}\ket{z_1}\ket{x_2}\ket{z_2} \cdots
\ket{x_d}\ket{z_d} \mapsto \ket{x_1}\ket{z_1 \oplus
f(x_1)}\ket{x_2}\ket{z_2 \oplus f(x_2)} \cdots
\ket{x_d}\ket{z_d \oplus f(x_d)}.\]
We are given $k$ distinct items $y_1, y_2, \ldots, y_k \in \{0,1\}^m$
in $k$ registers.

\item[Promise:] All the $y_i$'s are in the database $f$.

\item[Goal:] To devise a a quantum circuit with the minimum number of
applications of $T^{\tensor n}_f$ in order to determine the location
of each item.
\end{description}
}}
\caption{The problem}
\label{fig:original}
\end{figure}
In the following, we assume that $N$ is large and that $d$ and $k$ are much
smaller than $N$, say less than $\sqrt{N}$. Also, when we say that a
quantum algorithm solves a certain problem, we mean that it returns
the correct answer with probability at least $\frac{3}{4}$.

\section{Lower bound}

To present our lower bound argument, it will be convenient to
reformulate the problem slightly so that standard lower bound
techniques can be applied directly. For a function $f:\{0,1\}^n
\rightarrow \{0,1\}^m$, let $f^{\tensor d}:
\{0,1\}^{dn} \rightarrow \{0,1\}^{dm}$ be defined by 
\[ f^{\tensor d}(x_1,x_2,\ldots,x_d) =
(f(x_1),f(x_2),\ldots,f(x_d)).\]
Thus, $f^{\tensor d}$ is a database with $dn$-bit addresses and
$dm$-bit items, and we can associate as before the unitary
transformation 
$T_{f^{\tensor d}}$ with it acting on the $(dn+dm)$-qubit space. Also,
we have (with the natural reordering of coordinates) $T_{f^{\tensor
d}} = T_f^{\tensor d}$.  It is easy to see that if we have an
efficient solution to the problem in Figure~\ref{fig:original}, then
we have a solution with essentially the same complexity for the
following problem.
\begin{figure}
\framebox{\parbox[t]{5.5in}{
\begin{description}
\item[Input:] We are given a database $F:\{0,1\}^{dn}
\rightarrow \{0,1\}^{dm}$ such that $F= f^{\tensor d}$ for some (unique)
$f:\{0,1\}^n \rightarrow \{0,1\}^m$. We are given access to $F$ via
the unitary transformation $T_F$. We are given $k$ distinct items
$y_1, y_2, \ldots, y_k \in \{0,1\}^m$ in $k$ registers.

\item[Promise:] Either all $y_i$'s are in $f$ or exactly $k-1$ of them
are in $f$.

\item[Goal:] To devise a quantum circuit with the minimum number of
applications of $T_F$ in order to determine (with high probability) if
if all $y_i$'s are in $f$.
\end{description}
}}
\caption{The reformulated problem}
\label{fig:reformulated}
\end{figure}

With this formulation we can state our lower bound result.
\begin{theorem} \label{thm:lb}
Let $k\leq 2^{m-1}$. Any quantum circuit for solving the problem in 
Figure~\ref{fig:reformulated} requires 
\[ \Omega\left(\sqrt{\frac{Nk}{d \min\{d,k\}}}\right).\]
applications of the transformation $T_F$.
\end{theorem}

We will make use of the following special case of a result of
Ambainis~\cite{Ambainis} (see also Laplante and Magniez~\cite{LM}).
\begin{theorem} \label{thm:ambainis}
Let $G=(V_0,V_1,E)$ be a bipartite graph whose vertices are databases.
The edges of this database have labels. The edge $(F_0,F_1)$
connecting databases is labeled by all addresses $a$ such that $F_0(a)
\neq F_1(a)$. Let $\Delta_0$ be the minimum degree of a vertex in $V_0$ and
$\Delta_1$ be the minimum degree of a vertex in $V_1$. Let $\ell_0$ be the
maximum number of edges incident on a fixed vertex $F_0 \in V_0$ with
the same address. Similarly, let $\ell_1$ be the maximum number of edges
incident on a fixed vertex $F_1 \in V_1$ labeled with the same address.
Suppose there is a quantum circuit that returns the value $0$ with
high probability for databases in $V_0$ and returns the value $1$ with
high probability for databases in $V_1$. Then, this circuit
must contain
\[ \Omega\left(\sqrt{\frac{\Delta_0 \Delta_1}{\ell_0\ell_1}}\right)\]
applications of the unitary transform $T_F$.
\end{theorem}

With this, we are now ready to prove our lower bound.

\begin{proof} (of Theorem~\ref{thm:lb}) Fix any $k$ distinct items 
$y_1,y_2,\ldots,y_m \in \{0,1\}^m - \{0^m\}$. Let $V_0$ consist of all
databases $f_0^{\tensor d}$ where $f_0$ ranges over databases that contain
exactly $k-1$ of the $y_i$'s and the other $2^{n}-k+1$ locations
contain the item $0^n$. Thus, there are exactly, ${2^n \choose {k-1}}
k!$ vertices in $V_0$. Similarly, $V_1$ consists of databases of the
form $f_1^{\tensor d}$ where $f_1$ has all $y_i$'s and the other locations
contain $0^n$. The pair $(f_0^{\tensor d}, f_1^{\tensor d})$ is an
edge iff $f_0$ and $f_1$ differ in exactly one location. 

Clearly, the quantum circuit returns the answer $0$ with high
probability for databases in $V_0$ and returns the answer $1$ with
high probability for databases in $V_1$. To get our lower bound, it
remains only to compute the items $\Delta_0, \Delta_1, \ell_0$ and
$\ell_1$ for this graph. It is easy to check that $\Delta_0=N-k+1$
(there are these many locations in $[N]$ where we can introduce the
missing item) and $\Delta_1=k$ (there are these many ways to delete a
item). 

To determine $\ell_0$, fix a database $f_0^{\tensor d} \in V_0$ and a
location $(x_1,x_2,\ldots,x_d)$, suppose the missing item in $f_0$ is
$y_1$. Let the item stored at this location be
$(v_1,v_2,\ldots,v_d)$. In an adjacent database, $f_1^{\tensor d}$, if
the contents of this location are different, then one of the $v_i$'s
($d$ possibilities) must change from $0^n$ to $y_1$. It is easy to
verify that there are at most $d$ choices for $f_1^{\tensor d}$. 

To determine $\ell_0$ fix a database $f_1^{\tensor d} \in V_1$ and a
location $(x_1,x_2,\ldots,x_d)$. Let the item stored at this location
be $(v_1,v_2,\ldots,v_d)$.  In an adjacent database, $f_0^{\tensor
d}$, if the contents of this location are different, then one of the
$v_i$'s must change from being a $y_i$ and become $0^m$. It is easy to
verify that we have at most $\min\{d,k\}$ choices for $f_0^{\tensor
d}$. 

Our claim now follows immediately from Theorem~\ref{thm:ambainis}.
\qed
\end{proof}

\section{The algorithm}
In this section, we design a quantum algorithm with performance close
to the lower bound proved in the previous section.
\begin{lemma} \label{lm:Mark}
 Given one copy of the database of size $N$, and 
a promise that at most $t$ items out of $y_1,y_2,\ldots,y_k$ are in the
database, we can determine the location of all these $t$ items 
using $O(\sqrt{Nt})$ queries.
\end{lemma}
\begin{proof}
The algorithm of Figure~\ref{fig:repeated} makes a total of
$O(\sqrt{Nt})$ queries.
\begin{figure}
\framebox{\parbox[t]{5.5in}{
Let $Y=\{y_1,y_2,\ldots,y_k\}$.
Perform the following step for $i=1,2,\ldots,t$:
\paragraph{Step $i$:} Search for an item from $Y$
 in the database assuming that there are $t-i+1$ of them in the
 database. If an item $y$ is found, set $Y \leftarrow Y \setminus \{y\}$.
 This requires $O\left(\sqrt{\frac{N}{k-i+1}}\right)$ queries to the database.
}}

\caption{Algorithm for searching $k$ item using one database}
\label{fig:repeated}
\end{figure}
\end{proof}

We can now state our upper bound.

\begin{theorem} \label{thm:ub}
\begin{enumerate}
\item If $k\leq \sqrt{d}$, then the locations of all $k$ items can be
determined with $O\left(\sqrt{\frac{N}{d}}\right)$ parallel queries.

\item If $\sqrt{d} < k \leq d \lg d$, then the locations of all $k$
  items can be determined with
$$O\left(\sqrt{\frac{Nk \lg d}{d \min\{k,d\}}}\right)$$ parallel queries.

\item If $d\lg d < k$, then the location of all $k$ items can be
  determined with
$$O\left(\frac{\sqrt{Nk}}{d}\right)$$ parallel queries.

\end{enumerate}

\end{theorem}

\begin{proof}  If $d$ is a small constant, Lemma~\ref{lm:Mark} already
implies our theorem. So, we assume that $d$ is large.  The idea is to
ensure that different databases are used to search different parts of
the addresses. However, to balance the load on the different copies,
we assign random subsets of addresses to the different databases.
More formally, we randomly partition the address set $[N]$ into $d$
disjoint sets each of size $N/d$.  Then, we search these sets in
parallel dedicating one database for each. The probability that any
set has more than $t$ items is at most
\begin{equation}
{k \choose t} \left(\frac{1}{d}\right)^t. \label{max}
\end{equation}
We will choose the value of $t$ so that this quantity is much less than
$1/d$, so that with constant probability each set has at most $t$
items. Then, we can apply Lemma~\ref{lm:Mark} and obtain an algorithm
that makes $O(\sqrt{Nt/d})$ parallel queries, and with constant
probability determines the location of all items. We repeat the
algorithm several times to reduce the probability of error.

\paragraph{$k \leq \sqrt{d}$:}  Take $t=2$, and conclude from
(\ref{max}) that the probability that any one set has at least two
items is at most $1/(2d)$. Since, there are at most $d$ sets, 
with probability $\frac{1}{2}$ all sets have at most $1$ item.

\paragraph{$\sqrt{d} < k \leq d$:} In this case, take $t=5 \lg d$ and
conclude from (\ref{max}) that the probability that some set has at least $t$ elements
is at most $\frac{1}{d^3}$. Thus with probability at least
$\frac{1}{d}$ all sets have fewer than $5 \lg d$ elements.

\paragraph{$d < k \leq d\lg d$:} In this case, take $t = \frac{5k\lg
d}{d}$. The rest of the arguments is the same as before.

\paragraph{$k > d \log d$:}  In this case, take $t =
\frac{2k}{d}$. The rest of the argument is the same as before.

It thus follows that in all cases our claim holds.
\end{proof}

\subsection*{Acknowledgments} We thank Mark Heiligman for introducing
us to this problem and sharing with us the algorithm described in
Figure~\ref{fig:repeated}. This work was supported by 
NSA and ARO contract DAAG55-98-C-0040.

\end{document}